\documentclass{elsart}
\usepackage{epsfig}
\usepackage{amssymb}

\newtheorem{theorem}{Theorem}
\newtheorem{lemma}{Lemma}
\newtheorem{proposition}{Proposition}
\newtheorem{definition}{Definition}

\newcommand{\tV}{\tilde{V}}

\begin{document}

\begin{frontmatter}

\title{On existence of dark solitons in cubic-quintic nonlinear
Schr\"{o}dinger equation with a periodic potential}

\author{Pedro J. Torres}
\address{Departamento de Matem\'atica Aplicada, Universidad de Granada, 18071 Granada, Spain.}
\thanks{The work
of PJT was supported by D.G.I. MTM2005-03483, Ministerio de
Educaci\'on y Ciencia, Spain. The work of VVK was supported by the
Funda\c{c}\~ao para a Ci\^encia e a Tecnologia (Portugal) and
European program FEDER under the grant POCI/FIS/56237/2004.}

\author{Vladimir V. Konotop}
\address{Centro de F\'{\i}sica Te\'{o}rica e Computacional,
Universidade de Lisboa, Complexo Interdisciplinar, Av. Prof. Gama
Pinto 2, Lisbon 1649-003, Portugal and Departamento de
F\'{\i}sica, Universidade de Lisboa, Campo Grande, Ed. C8, Piso 6,
Lisboa 1749-016, Portugal}

\begin{abstract}
A proof of existence of stationary dark soliton solutions of the
cubic-quintic nonlinear Schr\"{o}dinger equation with a periodic
potential is given. It is based on the interpretation of the dark
soliton as a heteroclinic on the Poincar\'e map.
\end{abstract}

\journal{Comm.Math.Phys.}

\end{frontmatter}

\section{Introduction}

In the present paper we consider stationary solutions of the
cubic-quintic nonlinear Schr\"{o}dinger equation (CQNLS)
\begin{eqnarray}
\label{CQNLS} i \psi_t+\psi_{xx}+V(x)\psi-g_1|\psi|^2\psi-
g_2|\psi|^4\psi= 0
\end{eqnarray}
with an $L$-periodic symmetric potential: $V(x)=V(x+L)=V(-x)$ and
real constants $g_{1}$ and $g_2$. More specifically, we are
interested in solutions which allow the representation
$\psi(t,x)=e^{-i\omega t}\phi(x)$ where the function $\phi$ solves
the stationary equation
\begin{eqnarray}
\label{phi}
    \phi_{xx}+\tilde{V}(x)\phi-g_1\phi^3-g_2\phi^5=0
\end{eqnarray}
with $\tilde{V}(x)\equiv\omega+V(x)$,
 and which can be identified as {\em dark solitons} due to their
nonzero boundary conditions
\begin{eqnarray}
\label{boundary}
    \phi(x)\to  \phi_\pm (x)\quad \mbox{as}\quad x\to\infty
\end{eqnarray}
 with the functions $\phi_\pm (x)$ being real, sign definite, and
$L$-periodic solutions of (\ref{phi}). As it is clear, $\phi(x)$
acquires a zero value at some point of space, and therefore
without loss of generality can be searched real, what is taken
into account in the passage from (\ref{CQNLS}) to (\ref{phi}).

The model (\ref{CQNLS}) having a general character, it describes
weakly dispersive and weakly nonlinear wave processes, recently
attracted considerable attention in connection with its
application to the mean-field theory of Bose-Einstein condensates
\cite{Book}. In this context, $\psi(x)$ is a macroscopic wave
function, $|\psi(x)|^2$ described linear atomic density, and
$V(x)$ is an optical lattice created by standing laser beams. In
particular, existence of spatially localized pulses (also referred
to as bright solitons) has been recently addressed in Refs.
\cite{AS,AKP}. In Ref.~\cite{AKP} families of the solutions were
presented and significant differences in behavior of stationary
modes of the standard cubic nonlinear Schr\"{o}dinger equation
($g_1\neq 0$, $g_2=0$) and of the quintic nonlinear
Schr\"{o}dinger (QNLS) equation ($g_1= 0$, $g_2\neq 0$), have been
found. Dark solitons of the NLS equation with a periodic potential
have also been discussed in the small amplitude limit \cite{KS}
and for a general case it has been studied numerically in
Ref.~\cite{AKS} (see also \cite{BK} and references therein). The
approach developed in \cite{AKS} was based on the numerical study
of Poincar\'e map generated by Eq. (\ref{phi}) considered at
instants $nL$, the dark solitons appearing as {\em heteroclinics}
on the map. The aim of the present paper is to extend earlier
studies, providing for the first time a rigorous proof of the
existence of a dark soliton solution of Eq. (\ref{CQNLS}). From
the mathematical point of view, the strategy of proof combines in
a novel way several techniques from the classical theory of ODE's
(upper and lower solutions \cite{CH} and truncature arguments) and
dynamics of planar homeomorphisms (topological degree
\cite{Ma,Ortega} and free homeomorphisms \cite{B}).

We will restrict the consideration to the case $g_2>0$ only, what
rules out any possibility of blowing up solutions. Then without
loss of generality we can set $g_2=1$ through a rescalling, what
is done in what follows.

\section{Existence of one-signed periodic solutions}

We start with a proof of existence of one-signed periodic
solutions. To this end we impose the condition of boundness of
$V(x)$: $V_{min}\leq V(x) \leq V_{max}$, consider $\omega>-V_{min}$, and introduce the
notations $\lambda_1^2=\omega + V_{min}$ and  $\lambda_2^2=\omega + V_{max}$. As it is
clear $0<\lambda_1^2\leq\tV(x)\leq\lambda_2^2$. Next we consider
two stationary equations ($j=1,2$)
\begin{eqnarray}
    \phi_{j,xx}+\lambda_j^2\phi_j-g_1\phi_j^3-\phi_j^5=0
\end{eqnarray}
Treating these equations as dynamical systems, one easily finds
the (only) two nontrivial equilibria $\pm\rho_j$ where
\begin{eqnarray}
\label{rho}
    \rho_j=\sqrt{ \sqrt{g_1^2+4\lambda_j^2}-g_1}/\sqrt{2}
\end{eqnarray}
These are the hyperbolic points $+\rho_j$ and $-\rho_j$ which are connected by the
heteroclinic orbits, which explicit forms read
\begin{eqnarray}
\label{dark_j}
    \phi_j=\frac{\rho_j\alpha_j
\tanh(k_jx)}{\sqrt{\rho_j^2+\alpha_j^2-\rho_j^2\tanh^2(k_jx)}}
\end{eqnarray}
where
\begin{eqnarray}
    \alpha_j=\sqrt{2\rho_j^2+\frac 32 g_1},\quad\mbox{and}\quad
k_j=\rho_j\sqrt{\rho_j^2+\frac{g_1}{2} }
\end{eqnarray}

Let us call $\rho_{1,2}$ the positive equilibria. As it is clear
$\phi_1(0)=\phi_2(0)=0$ and $\phi_1(x)<\phi_2(x)$ for $x>0$.

At this point some considerations are required about the general
second order equation
\begin{eqnarray}
    \label{sec_order}
    \phi_{xx}=f(x,\phi)
\end{eqnarray}
with $f(x,\phi)$ continuous with respect to the both arguments and
$L$-periodic in $x$. The following definition is classical (see
for instance \cite{CH} and its references).

\begin{definition} A function $\alpha:[x_0,+\infty)\to\mathbb{R}$
such that $\alpha_{xx}(x)>f(x,\alpha)$ ($\alpha_{xx}(x)<f(x,\alpha)$) for all $ x>x_0$ is called
a {\em lower} ({\em upper}) solution of eq. (\ref{sec_order}).
\end{definition}

Now we can formulate
\begin{proposition}
\label{prop1} $\rho_1$ and $\rho_2$ are respectively lower and
upper solutions of equation (\ref{phi}). Hence here exists an
unstable $L$-periodic solution between them.
\end{proposition}

{\it Proof}:  Let us observe that for $j=1$
\begin{eqnarray}
\rho_{1,xx}+\tV(x)\rho_1-g_1\rho_1^3-\rho_1^5>\lambda_1^2\rho_1-g_1\rho_1^3-\rho_1^5=0
\end{eqnarray}
and similarly for $j=2$. Hence, $\rho_1<\rho_2$ are a couple of
well-ordered lower and upper solutions respectively, therefore
there exists a periodic solution between them \cite{CH}. Such
solution is unstable because the associated Brouwer index to the
Poincar\'e map is $-1$ (see for instance \cite{Ortega}). \qed

Therefore we have a positive $L$-periodic solution of Eq.
(\ref{phi}), we designate it as $\phi_+(x)$, such that
$\rho_1\leq\phi_+(x)\leq\rho_2$. By the symmetry of equation we
also have a negative solution $\phi_-(x)=-\phi_+(x)$.

To give an example of a periodic solution, we consider the
simplified QNLS model (\ref{CQNLS}) with $g_1=0$ and  with the
potential
\begin{eqnarray}
\label{V_example}
V(x)=\rho^4[2-k^2{\rm sn}^2(\rho^2 x,k)]^2
\end{eqnarray}
where sn$(x,k)$ is the
Jacobi elliptic function, $k\in[0,1]$ is the elliptic modulus, and $\rho>0$
(examples of the exact periodic solutions for the cubic nonlinear
Schr\"{o}dinger equation,  $g_2=0$, and with a specific potential,  were obtained
in~\cite{bronsk1,bronsk2}). The respective positive definite
solution reads
\begin{eqnarray}
\label{phi+}
\phi_+(x)=\rho\,{\rm dn}(\rho^2x,k)
\end{eqnarray}
 and it corresponds to the
frequency $\omega=\rho^4(k^2-3)$. To verify the stability of $\phi$ in the
sense of the dynamical system (\ref{phi}) we consider small deviation
$\psi(x)=\phi(x)-\phi_+(x)$ at $x\to\infty$, whose dynamics in the leading order is
governed by the equation
\begin{eqnarray}
\label{Hill}
    \psi_{xx}-U(x,k)\psi=0
\end{eqnarray}
with
\begin{eqnarray}
\label{u}
    U(x,k)=\rho^4(4-k^2-6k^2{\rm sn}(\rho^2x,k)^2+4k^4{\rm sn}(\rho^2x,k)^4)\geq
    \nonumber \\
    \geq
\rho^4(\frac 74-k^2)>0
\end{eqnarray}
Thus, the obtained function $\phi_+(x)$ is a hyperbolic periodic
solution of Eq. (\ref{phi}).

Considering now $\psi_+(x,t)=\phi_+(x)\exp(i(3-k^2)t)$ as a
solution of Eq. (\ref{CQNLS}), performing the stability analysis
as in Ref.~\cite{bronsk1}, only slightly modified due to presence
of quintic nonlinearity,  and taking into account that
$\phi_+(x)>0$ one verifies that $\psi_+(x,t)$ is linearly stable
in the sense of the evolution problem (\ref{CQNLS}).

More sophisticated models allowing exact sign definite periodic
solutions can be constructed using a kind of "inverse engineering"
(i.e. by obtaining potentials starting with given periodic
solutions) as it is explained in~\cite{BK}.

\section{Existence of a dark soliton}

In this section we prove the existence of a heteroclinic orbit
connecting the periodic solutions $\phi_{-}$ and $\phi_{+}$. A
battery of preparatory lemmas are necessary.

\begin{lemma}\label{lemma1} If $\phi:[x_0,+\infty)\to \mathbb{R}$ is a bounded
solution of eq. (\ref{sec_order}), then the derivative $\phi_x$ is
also bounded in $[x_0,+\infty)$.
\end{lemma}

{\it Proof}: By the hypothesis $|\phi(x)|<M$, where $M$ is a
constant, for all $x\geq x_0$. Then, by the mean value theorem
there exits $x_n\in(nL,(n+1)L)$ such that
$\phi((n+1)L)-\phi(nL)=\phi_x(x_n)L$. From here $\phi_x(x_n)<2M/L$
for all $n$. Applying the mean value theorem one more time one
obtains
\begin{eqnarray}
    |\phi_x(x)-\phi_x(x_n)|<L\max_{|\phi|\leq M}|f(x,\phi)|,\quad\forall
x\in(nL,(n+1)L)
\end{eqnarray}
because $|\phi_{xx}(x)|<\max_{|\phi|\leq M}|f(x,\phi)|$. Then
\begin{eqnarray}
    |\phi_x(x)|<L\max_{|\phi|\leq M}|f(x,\phi)|+\frac{2M}{L},\quad
\forall x\geq x_0.
\end{eqnarray}   \qed

The following lemma is a key ingredient in our main result.

\begin{lemma}
\label{lemma_CT} \cite{CT} Let $P:\mathbb{R}^2\to\mathbb{R}^2$ be
an orientation preserving homeomorphism with a unique fixed point
$p_L$ such that $\gamma\{I-P,p_L\}\neq 1$. Then for any
$p_0\in\mathbb{R}^2$ one of the following possibilities holds

i) $P^n(p_0)\to p_L$ as $n\to+\infty$

ii) $\|P^n(p_0)\|\to \infty$ as $n\to+\infty$
\end{lemma}

Here, $\gamma\{I-P,p_0\}$ is the local index associated to the
Brouwer degree of $p_0$ as a fixed point of the homeomorphism $P$.
The proof relies in a basic property of free homeomorphisms
exposed in \cite{B}, namely the $\omega$-limit set of a given
point has to be a connected set of the fixed point set.

\begin{lemma}
If $f(x,y)$ is strictly increasing in $y$, there exists a most one
$L$-periodic solution of (\ref{sec_order}).
\end{lemma}

{\it Proof}: By contradiction, let us assume that $y_1, y_2$ are
two different $L$-periodic solutions of (\ref{sec_order}). First,
let us suppose that $y_1, y_2$ intersect among themselves, that
is, there should be $t_1,t_2$ such that $z(t)=y_1(t)- y_2(t)$
verifies $z(t_0)=0=z(t_1)$ and $z(t)>0$ for $t\in(t_1,t_2)$.
However, by substracting the corresponding equations and using
that $f$ is strictly increasing, we get that $z$ should be convex
in $(t_1,t_2)$, which is a contradiction. Therefore, $y_1, y_2$ do
not intersect and we assume without loss of generality that
$y_1(t)>y_2(t)$ for all $t$. Again, $z$ should be convex in the
whole real line, but this is impossible because it is
periodic.\qed

With the help of these previous lemmas we are able to prove an
abstract convergence result.

\begin{theorem}
\label{theor1} Let $\phi:[x_0,+\infty)\to\mathbb{R}$ a bounded
solution of (\ref{sec_order}). Let us assume that
\begin{eqnarray}
    \min_{x\in[0,L]\atop y\in[\inf_{x\geq x_0}\phi(x), \sup_{x\geq
x_0}\phi(x)]}\frac{\partial f(x,y)}{\partial y}>0
\end{eqnarray}
Then there exists an $L$-periodic solution $\varphi(x)$ such that
\begin{eqnarray}
    \lim_{x\to+\infty}(|\phi(x)-\varphi(x)|+|\phi_x(x)-\varphi_x(x)|)=0
\end{eqnarray}
\end{theorem}

{\it Proof}: Let us define $m=\inf_{x\geq x_0}\phi(x)$ and
$M=\sup_{x\geq x_0}\phi(x)$, as well as the truncated function
\begin{eqnarray}
    \tilde{f}(x,y)=\left\{
    \begin{array}{l}
    f(x,y),\qquad \forall y\in [m,M]
    \\
    f(x,M)+f_y(x,M)(y-M),\qquad \forall y>M
    \\
    f(x,m)+f_y(x,M)(y-m),\qquad \forall y<m
    \end{array}
    \right.
\end{eqnarray}
$\tilde{f}$ is strictly increasing in $y$. Note also that $\phi$
is a solution of the truncated equation
\begin{eqnarray}
    \label{sec_order_trunc}
    \phi_{xx}=\tilde{f}(x,\phi)
\end{eqnarray}
Obviously, $\lim_{y\to\pm\infty}\tilde{f}(x,y)=\pm\infty$
uniformly in $x$. Hence, there exist constants $\alpha$ and
$\beta$, such that $\alpha<\beta$ and
$\tilde{f}(x,\alpha)<0<\tilde{f}(x,\beta)$ for all $x$. Such
$\alpha$ and $\beta$ is a well-ordered coupled $L$-periodic lower
and upper solutions, so there exists an $L$-periodic solution of
(\ref{sec_order_trunc}) with index $-1$. This solution is unique
by the previous lemma. Then, by Lemma~\ref{lemma_CT}, $\phi(x)$
must converge to $\varphi(x)$ since Lemma \ref{lemma1} excludes
$ii)$. As $\varphi(x)\in[m,M]$, it is a solution of
(\ref{sec_order}). \qed

\begin{theorem}
\label{theor2}
 Let us consider bounded functions
$\alpha,\,\beta:[x_0,+\infty)\to\mathbb{R}$ verifying

 1) $\alpha(x)<\beta(x)$, $\forall x>x_0$

 2) $\alpha_{xx}(x)>f(x,\alpha)$ and $\beta_{xx}(x)>f(x,\beta)$,
$\forall x>x_0$

Then there exists a solution $\phi(x)$ of (\ref{sec_order}) such
that
\begin{eqnarray}
\alpha (x)< \phi(x) < \beta (x)
\end{eqnarray}
If moreover, there exists $x$ such that

3) $\displaystyle{\min_{x\in[0,L]\atop y\in[\inf_{x\geq
x_0}\alpha(x), \sup_{x\geq x_0}\beta(x)]}\frac{\partial
f(x,y)}{\partial y}>0}$

then there exists an $L$-periodic solution $\varphi(x)$ such that
\begin{eqnarray}
    \lim_{x\to+\infty}(|\phi(x)-\varphi(x)|+|\phi_x(x)-\varphi_x(x)|)=0
\end{eqnarray}
Besides, $\varphi(x)$ is the unique $L$-periodic solution in the
interval  $[\inf_{x\geq x_0}\alpha(x), \sup_{x\geq x_0}\beta(x)]$.
\end{theorem}

{\it Proof}: The first assertion is a classical result due to
Opial \cite{O}. The second conclusion is a corollary of
Theorem~\ref{theor1}. $\Box$

In order to apply the above results to our model (\ref{CQNLS}), we
observe that i) now $f(x,y)\equiv y^5+g_1y^3-\tV(x)y$, ii) due to
parity of the potential one can consider $x\geq 0$ and extend the
obtained solution $\phi(x)$ as an odd function to $x\leq 0$, iii)
The functions $\phi_{1,2}(x)$ given by (\ref{dark_j}) satisfy the
conditions 1) and 2) of the Theorem~\ref{theor2} where
$\alpha(x)\equiv\phi_1(x)$ and $\beta(x)\equiv\phi_2(x)$. Hence,
in order to prove that there exists a solution $\phi(x)$ of
(\ref{CQNLS}) converging to $\phi_\pm(x)$, found in
Proposition~\ref{prop1}, as $x\to\pm\infty$, one has to verify the
condition 3) of Theorem~\ref{theor2}.

As $x_0$ can be taken arbitrarily large, this last condition is
equivalent to
\begin{eqnarray}
\label{cond2}
    \min_{y\in[\rho_1,\rho_2]\atop
x\in[0,L]}\left\{5y^4+3g_1y^2-\tV(x)\right\}>0
\end{eqnarray}
Starting with the case $g_1\geq 0$ we observe that (\ref{cond2})
is now equivalent to $5\rho_1^4+3g_1\rho_1^2-\lambda_2^2>0$. The
straightforward analysis of this last inequality, which takes into
account the link (\ref{rho}), the definition of $\lambda_{1,2}$,
the requirement $\omega>-V_{min}$  necessary for $\lambda_1^2>0$,
gives the following estimate for the frequency
\begin{eqnarray}
\label{omega1}
    \omega-g_1\sqrt{g_1^2+4\omega+4V_{min}}>V_{max}-5V_{min}
\end{eqnarray}
Thus (\ref{omega1}) is a sufficient condition for the existence of
dark solitons at non-negative $g_1$.

Considering now $g_1<0$ the constrain (\ref{cond2}) is reduced to
$5\rho_1^4-3|g_1|\rho_2^2-\lambda_2^2>0$ and subsequently to the
following inequality constrain to the frequency
\begin{eqnarray}
8\omega+2g_1^2+10V_{min}-2V_{max}-5g_1\sqrt{4\omega+g_1^2+4V_{min}}
\nonumber \\
+3g_1\sqrt{4\omega+g_1^2+4V_{max}}>0
\end{eqnarray}
which must be satisfied simultaneously with $\omega>-V_{min}$.

In both cases considered before, the conclusion is that there
exists an explicitly computable $\omega_0$ such that the CQNLS has
a different dark soliton for any $\omega>\omega_0$.

\section{A concluding example}

\begin{figure*}
\epsfig{file=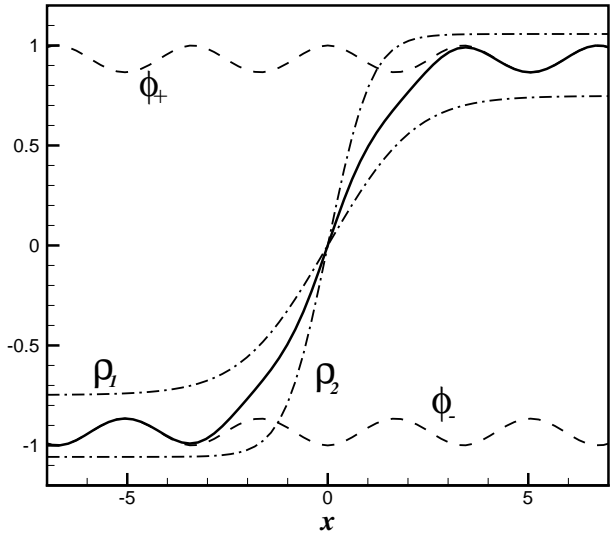,width=10cm} \caption{The dark soliton --€
heteroclinic (solid line), lower and upper solutions $\rho_{1,2}$
(dashed-dotted line), and the hyperbolic periodic solutions
$\phi_\pm(x)$ given by (\ref{phi+}), for the parameters $k=0.5$
and $\rho=1$} \label{figone}
\end{figure*}

As a concluding remark we consider an example illustrating a dark soltion, as well as other concepts introduced in the paper. To this end we
recall the potential (\ref{V_example}) and construct a dark soliton which tends to $\phi_+$ given by (\ref{phi+}) (we thus consider now
$g_1=0$). This cannot be done analytically, and that is why we employ numerics. An  example is shown in Fig.~\ref{figone}.

For numerical obtaining the dark soliton, we have use shooting
method. To this end we observed that (\ref{Hill}) is a Hill's
equation and thus taking into account (\ref{u}), from Floquet's
theorem one we have that $\psi(x)=P(x)\exp(-\alpha x)$ where
$P(x)=P(x+2{\rm K}(k)/\rho^2)$ is a periodic function and ${\rm
K}(k)$ is a complete elliptic integral of the first kind. For
given parameters $k$ and $\rho$ one can easily compute the
respective Floquet's exponent $\alpha$. In particular for our
choice of $k=0.5$ and $\rho=1$, $\alpha \approx 2.014298$.  Thus,
starting with the points $x_{ini}=2{\rm K}(k)/\rho^2$ where $n$ is
an integer, where $\phi(x_{ini})=\phi_+(0)-C$ and
$\phi_x(x_{ini})=-C\alpha$, and by varying the parameter $C$ one
can meet the condition $\phi(0)=0$. An example of implementation
of this procedure is shown in Fig.~\ref{figone}.

\end{document}